**Research Article**

André C. A. Siqueira[1,*], Guillermo Palacios[1], Jessica E. Q. Bautista[1], Anderson M. Amaral[1], Albert S. Reyna[2], Edilson L. Falcão-Filho[1], and Cid B. de Araújo[1,*]

1 Departamento de Física, Universidade Federal de Pernambuco, Recife 50670-901, PE, Brazil

2 Programa de Pós-Graduação em Engenharia Física, Unidade Acadêmica do Cabo de Santo Agostinho, Universidade Federal Rural de Pernambuco, Cabo de Santo Agostinho 54518-430, PE, Brazil

# Observation of Replica Symmetry Breaking in Filamentation and Multifilamentation

**Abstract:** We report the experimental observation and characterization of Replica Symmetry Breaking (RSB) manifestation while analyzing the transverse intensity profile of pulses in filamentation experiments using sapphire crystal and distilled water, excited by a femtosecond laser centered at 800 nm. The RSB arises from the competition between self-focusing and plasma defocusing, subject to local fluctuations in the nonlinear refractive index generated by plasma via multiphoton excitation, which subsequently promotes frustration among modes. Our results confirm the existence of glassy-like photonic states not only in multifilamentation, as previously reported [*W. Ettoumi, J. Kasparian, and J. Wolf, "Spin-glass model governs laser multiple filamentation," Phys. Rev. Lett., vol. 115, no. 3, pp. 033902, 2015*], but also in the generation of a single filament and in filamentation accompanied by conical emission. These findings improve the understanding of statistical nonlinear optics by establishing connections with magnetism and highlighting the glassy-like behavior of light in the context of ultrafast optical phenomena.

## 1 Introduction

Ultrashort pulse filamentation represents an ultimate regime of light-matter interaction, characterized by an intense core that continuously exchanges energy with the surrounding photon reservoir. This dynamic structure can propagate over distances far exceeding the typical Rayleigh length while maintaining a narrow beam size without any external guiding mechanism [1-5]. This behaviour is governed by the balance between Kerr self-focusing and plasma defocusing, which is mediated by multiphoton excitation (MPE) and influenced by space-time effects such as self-phase modulation (SPM), diffraction, and dispersion [1-5].

Due to the universality of filamentation, which occurs in gases, transparent solids, and liquids, it has attracted significant interest within the ultrafast phenomena community. This interest has spurred numerous experimental and theoretical studies under various conditions. For instance, filamentation is of key importance for production of attosecond pulses during the step of high-harmonic generation [6-9]. In addition, in many of the few-optical-cycle pulse sources, filamentation is the process exploited in order to obtain the coherent broadband spectrum required to support such pulses [10-13]. In atmospheric analysis and remote sensing, filamentation is employed to enhance, over long distances, the detection and measurement of numerous atmospheric components and phenomena [14-19]. Laser micromachining, when incorporates filamentation as part of the process, enhances the overall precision of remote drilling and cutting of metals and biological materials [20,21]. These diverse applications underscore the importance of understanding filamentation in order to drive continued research and innovation in laser based technologies [22].

In transparent media, filamentation at powers higher than the threshold generally results in a rainbow-like angular emission known as conical emission (CE). Various mechanisms have been proposed to explain this phenomenon, including Cerenkov radiation [23,24], self-phase modulation (SPM) [25,26], four-wave mixing [27,28], and X-waves [29-31]. However, the statistical behavior of the transversal intensity profile of both core filaments and conical emission remains incompletely understood, despite the rogue-like events observed in the supercontinuum spectrum obtained from filamentation [32-34]. Gaining a deeper understanding is crucial due to the broad range of applications, which also opens possibilities for drawing analogies with other physical phenomena

At input powers significantly above the threshold for observing filamentation and conical emission, small perturbations (noise) in the beam profile or local fluctuations in the refractive index can cause the beam to break up into multiple 'hot spots' that act as nuclei for filaments, which then distribute stochastically. This phenomenon, known as multifilamentation, involves interactions between multiple filaments that are dependent on the phase differences between them [35-38].

Recently, filamentation has been a subject of study within statistical nonlinear optics, aiming to demonstrate that this critical phenomenon can be associated with phase transitions [37] and analogies across different physical systems that transcend nonlinear optics. It has been shown that the onset of multiple filamentation patterns can exhibit behaviors analogous to either solid or liquid phases, depending on the interactions between neighboring filaments [38]. Interestingly, multifilamentation has also been described through a lattice spin model with glassy-like dynamics [39], analogous to magnetism. In this context, unlike passive mode-locked lasers, which exhibit typical behavior of ferromagnetic-like systems (where the spins align in the same direction without frustration events), interacting filaments fail to align in a spatially regular configuration due to frustrated interactions.

In the context of detecting photonic glassy states, Replica Symmetry Breaking (RSB) provides a fundamental theoretical framework. A key tool in this analysis is the Parisi overlap distribution (or Parisi's probability distribution, PPD), $P(q)$, which characterizes the correlation among independent replicas of the system analyzed under the same experimental conditions [40-43]. Random lasers represent the first phenomenon in which glassy behavior was demonstrated through RSB. In this case, below the RL threshold, the fluorescence exhibits uncorrelated fluctuations analogous to paramagnetic-like behavior, while starting from the RL threshold, the frustrated mode competition exhibits RSB. Specifically, the PPD changes from a $P(q)$ centered at 0 ($|q_{max}| = 0$) to a bimodal shape ($|q_{max}| \approx 1$), indicating that fluctuations among spectral replicas begin to exhibit both correlation and anti-correlation, e therefore, photonic glassy-like behavior [44-51].

[*]**Corresponding author: André C. A. Siqueira**, E-mail: andrechaves.physics@gmail.com; and **Cid B. de Araújo,** E-mail: cid.araujo@ufpe.br

In addition to random lasers, RSB has also been observed in optical phase transitions in other photonic systems. For instance, it occurs in wave propagation subject to the interplay between disorder and nonlinearity [52]. In Q-switched Nd:YAG system, unlike passive mode-locked lasers, the behavior is similar to a random bond ferromagnet, where most spins align in the same direction [53]. In this scenario, the modes display some degree of phase coherence, and RSB arises from the random activation of specific coherent mode subsets after each excitation pulse, which frustrates the other modes.

More recently, RSB was also reported at the transition state between continuous wave (CW) and passive mode-locking in ytterbium-based mode-locked fiber lasers. Here, nonlinearity and disorder arising from the competition for gain result in RSB, revealing a glassy-like state between the paramagnetic-like regime (CW) and the ferromagnetic-like regime (passive mode-locked) [54]. In addition to classical systems, RSB has also been reported in the nontrivial quantum dynamics of a confocal cavity QED system. In this case, it was demonstrated that the quantum trajectories of entangled spins exhibit frustration in a spin network consisting of up to 15 spin-1/2 particles [55].

Regarding filamentation, local fluctuations in the refractive index induced by the generated plasma can cause certain regions of the pulse to experience varying degrees of self-phase modulation (SPM) or multiphoton excitation (MPE). This creates a competitive environment where different nonlinear effects, influenced by local refractive index disorder, interact in a way that not all interactions can be simultaneously satisfied. This frustration among modes opens the possibility for RSB.

In this paper, within the framework of Parisi's Probability Distribution (PPD), we analyze the statistical regimes of the transverse intensity profile of the pulse during the optical transition to filamentation, filamentation accompanied by conical emission, and multifilamentation in two different transparent media: a 5 mm-thick sapphire crystal and a 30 mm-thick cuvette filled with distilled water. Our findings reveal that before the onset of filamentation, only replica-symmetric behavior is observed, indicating a ferromagnetic-like statistical regime (passive mode-locked laser). However, RSB is observed in three distinct scenarios: around the filamentation threshold, during saturated filamentation accompanied by conical emission, and in multifilamentation. Our results highlight the glassy nature of the nonlinear competition between self-focusing and plasma defocusing, sustained by the frustration among modes.

The paper is organized as follows: Section 2 describes theoretical details and an introduction to the analysis procedure used to study the statistical behavior of filamentation. Section 3 presents the experimental results and discusses the Replica Symmetry Breaking phenomenon observed. Finally, Section 4 summarizes the results.

## 2 Theoretical framework

In exploring the theoretical aspects of filamentation, a natural starting point is to describe the system through a Lagrangian formalism. This approach allows us to systematically derive the underlying dynamics of the nonlinear Schrödinger equation (NLSE) and the corresponding Hamiltonian density, capturing the key physical processes governing the evolution of intense light filaments. By carefully constructing the Lagrangian, we can incorporate effects such as Kerr nonlinearity, plasma generation, and multiphoton excitation, all of these terms play crucial roles in the filamentation process.

Consider the Lagrangian density for the electric field envelope $A(x,y,z,t) = A$ and its complex conjugate $A^*(x,y,z,t) = A^*$, expressed in the following form [56-58]:

$$\mathcal{L} = \left(A^*\frac{\partial A}{\partial z} - A\frac{\partial A^*}{\partial z}\right) + a\left(\frac{\partial A}{\partial x}\frac{\partial A^*}{\partial x} + \frac{\partial A}{\partial y}\frac{\partial A^*}{\partial y}\right) + b\frac{\partial A}{\partial t}\frac{\partial A^*}{\partial t} - c|A|^4 + d|A|^2 + \frac{e}{K}|A|^{2K}, \tag{1}$$

the coefficients $a, b, c, d$ and $e$ in Eq. (1) are:

$$a = \frac{i}{k}, \tag{2}$$

$$b = ik'', \tag{3}$$

$$c = ikn_2, \tag{4}$$

$$d = \sigma(1 + i\omega_0\tau_c)\rho(t), \tag{5}$$

$$e = \beta_K\left(1 - \frac{\rho(t)}{\rho_{at}}\right), \tag{6}$$

where, $k$ is the wavenumber, $k''$ is the group-velocity dispersion, $n_2$ is the nonlinear refractive index, $\sigma$ is the cross section for inverse Bremsstrahlung, $\tau_c$ is the electron collision time, $\rho(t)$ corresponds to the time-dependent electron density excited through the absorption of $K \geq 1$ photons (MPE), $\beta_K$ is the multiphoton excitation coefficient, and $\rho_{at}$ denotes the density of neutral atoms. Applying Eq.(1) into the Euler-Lagrange equation:

$$\frac{\partial \mathcal{L}}{\partial A^*} - \frac{d}{dz}\left[\frac{\partial \mathcal{L}}{\partial\left(\frac{\partial A^*}{\partial z}\right)}\right] - \frac{d}{dt}\left[\frac{\partial \mathcal{L}}{\partial\left(\frac{\partial A^*}{\partial t}\right)}\right] - \frac{d}{dx}\left[\frac{\partial \mathcal{L}}{\partial\left(\frac{\partial A^*}{\partial x}\right)}\right] - \frac{d}{dy}\left[\frac{\partial \mathcal{L}}{\partial\left(\frac{\partial A^*}{\partial y}\right)}\right] = 0,$$

the spatiotemporal NLSE, a widely recognized equation for describing the propagation of an optical pulse while incorporating both spatial and temporal effects, is derived as follows [5]:

$$\frac{\partial A}{\partial z} = \frac{a}{2}\left(\frac{\partial^2 A}{\partial x^2} + \frac{\partial^2 A}{\partial y^2}\right) + \frac{b}{2}\frac{\partial^2 A}{\partial t^2} + c|A|^2 A - \frac{d}{2}A - \frac{e}{2}|A|^{2K-2}A. \tag{7}$$

The first and second terms in Eq. (7) denote the diffraction and temporal dispersion of the pulse, respectively. The subsequent terms represent effects arising from the nonlinear interaction between light and matter, with the third term corresponding, respectively, to SPM (or Kerr effect), while the fourth and fifth terms account for plasma defocusing (and absorption) and multiphoton excitation (MPE).

In the context of filamentation in sapphire crystal and distilled water using an input pulse with a central wavelength of $\lambda_0 = 800\, nm$, the positive $n_2$ induces self-focusing, which amplifies the pulse's peak intensity and contributes to the generation of new frequencies through SPM. Concurrently, the normal dispersion regime ($k'' > 0$) mitigates these processes via temporal broadening, which reduces the pulse peak intensity and smoothes the temporal derivative of the pulse profile. Similarly, diffraction, which is responsible for the beam divergence, also contributes to decrease the peak intensity. However, under nonlinear conditions, eventually, the diffraction effect may be overcome since that the input pulse has enough peak power in order to induce strong self-focusing. As a result, the energy reservoir of the beam continues to drive the pulse's peak intensity higher through self-focusing until plasma is generated via multiphoton excitation (MPE), thereby preventing the catastrophic collapse of the pulse. This configuration occurs when [4]:

$$n_2 I = \frac{2\pi e^2 N_e}{n_0 m_e \omega_0}.$$

In general, in transparent media, when the free-electron density ($N_e$) reaches approximately $10^{17}$ - $10^{18}$ $cm^{-3}$, the plasma generated via MPE counteracts self-focusing by introducing a negative contribution to the refractive index. This causes the beam to achieve peak intensities ($I_{stop}$) on the order of a few $TW/cm^2$, depending on its bandgap energy [4].

As is known, the filamentation phenomenon is characterized by maintaining a narrow beam size over distances longer than the typical Rayleigh length without any external guiding mechanism, potentially promoting the generation of broadband spectra (supercontinuum generation) and conical emission across multiple cycles of self-focusing and defocusing [5]. In this regime, the generation of new frequencies is driven by contributions from SPM and plasma ($N_e$) [5,59]:

$$\omega_{inst} = \omega_0 - \frac{\omega_0 L n_2}{c}\frac{dI(x,y,t)}{dt} + \frac{2\pi L e^2}{c m_e \omega_0}\frac{dN_e(x,y,t)}{dt}, \quad (8)$$

where $L$ is the sample length and $c$ is the speed of light.

As the beam undergoes cycles of self-focusing and plasma defocusing, different portions of the pulse interact with the medium in varying states. For instance, the leading edge of the pulse always interacts with the medium in its fundamental state, while other parts of the pulse, particularly those following the peak, interact with both the dielectric medium and the generated plasma. In this complex process, the nonlinear dynamics of filamentation tend to drive the peak intensity toward $I_{stop}$, with the filamentation intensity acting as an attractor. Thus, the beam remains in the filamentation state until energy dissipation and/or temporal broadening no longer support the self-focusing effect.

Applying the Lagrangian density (Eq. 1) to the Legendre transformation:

$$\mathcal{H} = p\frac{\partial A}{\partial z} + p^*\frac{\partial A^*}{\partial z} - \mathcal{L},$$

where $p = \partial\mathcal{L}/\partial(\partial_z A)$ and $p^* = \partial\mathcal{L}/\partial(\partial_z A^*)$, the Hamiltonian density for filamentation is obtained:

$$\mathcal{H}_{Fil} = -a\left(\frac{\partial A}{\partial x}\frac{\partial A^*}{\partial x} + \frac{\partial A}{\partial y}\frac{\partial A^*}{\partial y}\right) - b\frac{\partial A}{\partial t}\frac{\partial A^*}{\partial t} + c|A|^4 - \\ - d|A|^2 - \frac{e}{K}|A|^{2K},$$

with $a, b, c, d$, and, $e$ given by Eqs. (2-6), the electric field envelope can described as a mode expansion, expressed as $A(x,y,z,t) = \sum_n A_n(x,y,z,t)$. From this theoretical framework, this mode expansion offers a convenient method for representing the temporal and spatial profile of an optical beam, where each mode contributes to the overall field envelope. Thus, given a certain propagation distance $z_0$, at any specific point in the beam $(x_0, y_0, t_0)$, the field envelope is described by the contribution of multiple modes, each with its own amplitude and phase. Consequently, the Hamiltonian density assumes a discrete form:

$$\mathcal{H}_{Fil} = \mathcal{H}_{Linear} + c\sum_{n,m,i,j} A_n A_m^* A_i A_j^* - d\sum_{n,m} A_n A_m^* \\ - \frac{e}{K}\sum_{\substack{n_1,n_2,\ldots,n_K \\ m_1,m_2,\ldots,m_K}} A_{n_1}A_{m_1}^* \ldots A_{n_k}A_{m_k}^*. \quad (9)$$

with each other through linear and nonlinear terms, and their interactions govern the filamentation process. The linear part of the Hamiltonian density ($\mathcal{H}_{Linear}$) represents the mode interactions through the transfer of energy between spatial modes (diffraction) and temporal modes (dispersion). These effects can be written as:

$$\mathcal{H}_{Linear} = -a(\psi_x + \psi_y) - b\psi_t,$$

where $\psi_\eta$ (with $\eta = x, y, t$) represents the contributions of the modes to the diffraction or dispersion. Specifically, these terms involve derivatives of the mode amplitudes with respect to the spatial and temporal coordinates:

$$\psi_\eta = \sum_{n,m}\frac{\partial A_n}{\partial\eta}\frac{\partial A_m^*}{\partial\eta}.$$

The Kerr effect (and SPM) introduces a nonlinear interaction where the intensity of the electric field modulates the phase of the modes. This is represented by a four-mode mixing term in the Hamiltonian density in Eq.(9). The interaction of four modes causes nonlinear phase shifts that depend on the amplitudes and phases of the modes. This coupling leads to self-focusing and spectral broadening (in both temporal and spatial domains), where the pulse intensity increases due to constructive interference, counteracting diffraction.

As the intensity increases, MPE and plasma generation become significant, counteracting the self-focusing effect and generating blue-shifted spectral components (Eq. (8)). In MPE, the nonlinear process requires multiple photons to interact simultaneously with the medium to excite or ionize atoms, involving the summation over the interaction of $2K$ modes: $K$ components representing absorption and $K$ components representing the conjugate emission. The coherent nature of this interaction is captured by the phases of the modes aligning in a specific way for the MPE process to occur effectively. On the other hand, plasma generation introduces a phase shift in the modes, creating local regions where the refractive index is reduced due to the presence of free electrons, leading to the aforementioned plasma defocusing and absorption. This effect is represented in the Hamiltonian density (Eq. (9)) by the interaction between two modes. As the pulse propagates through these regions, different parts of the pulse experience a kind of varying phase shifts, which can contribute to the degradation of the pulse's coherence.

By applying the mode expansion formalism to the Hamiltonian density, it has become clearer how the various phenomena involved interact from the perspective of the electric field envelope's components for each propagation distance $z$ in the $(x,y,t)$ coordinates. Thus, throughout the filamentation process, Eq. (9) provides valuable insight into how the different effects influence one another, suggesting possible correlations and anti-correlations due to the interplay among the involved effects. To obtain the system's Hamiltonian, the integration is carried out over the entire spatial and temporal volume, represented by $H = \int \mathcal{H}_{Fil} dxdydtdz$. This approach captures the total energy and dynamic behavior of the system as the pulse propagates through space and time, while the Hamiltonian density captures the local dynamics of the filamentation.

During filamentation, local fluctuations in the refractive index due to the generated plasma can induce regions where the pulse may experience more or less SPM or MPE, thereby creating a competitive scenario between different nonlinear effects influenced by local refractive index disorder. Under these circumstances, there is frustration among modes, as all interactions cannot be satisfied simultaneously, which opens possibility for RSB. In this context, it is natural to consider an additional disorder term in $a, b, c, d$, and, $e$ in the Hamiltonian density in Eq. (9). At this point, it is important to emphasize the mathematical parallels between the filamentation Hamiltonian (Eq. (9)) and the glassy Hamiltonian (see [41]), as both feature interaction terms involving two-mode and four-mode couplings of the field. This structural similarity indicates a potential deeper link between the nonlinear dynamics of optical filamentation and the statistical mechanics of disordered systems, further reinforcing the analogy with glassy states.

To investigate the statistical regime of filamentation within the framework of Parisi's probability distribution, we analyzed the distribution of the overlap parameter ($q_{\alpha\beta}$), which quantifies the correlation of intensity fluctuations in the transverse profiles across different replicas. The analyzed images have a resolution of 240 × 240 pixels, and the $q_{\alpha\beta}$ values were calculated using the following expression [60]:

$$q_{\alpha\beta} = \frac{\sum_{x_i}\sum_{x_j}\Delta_\alpha(x_i,x_j)\Delta_\beta(x_i,x_j)}{\sqrt{\sum_{x_i}\sum_{x_j}\left(\Delta_\alpha(x_i,x_j)\right)^2}\sqrt{\sum_{x_i}\sum_{x_j}\left(\Delta_\beta(x_i,x_j)\right)^2}}, \quad (10)$$

where $\alpha$ and $\beta = 1,2,3,\ldots,N$ represent the replica indices of the N-th replica, and $x_i$ and $x_j$ are the pixel indices corresponding to the images recorded by the CCD camera. The term $\Delta_\alpha(x_i,x_j) = I_\alpha(x_i,x_j) - \langle I(x_i,x_j)\rangle$ represents the intensity fluctuation at the pixel with coordinates $(x_i,x_j)$ for replica $\alpha$ (or $\beta$ for $\Delta_\beta(x_i,x_j)$), where $\langle I(x_i,x_j)\rangle = \frac{1}{N}\sum_\alpha I_\alpha(x_i,x_j)$ is the average intensity calculated over all replicas.

In addition to directly analyzing the intensity profiles in the position domain, we applied a two-dimensional Fourier transform to the images to examine Parisi's distribution of the spatial frequency profiles using Eq.(10). This enabled us to quantify the correlation of intensity fluctuations in the spatial frequency domain (Fourier space) across different replicas, offering a complementary perspective for analyzing the photonic glassy behavior of filamentation and multifilamentation.

## 3 Experimental results

The experiment was conducted using a regenerative Ti:sapphire amplifier that delivered linearly polarized optical pulses with central wavelength $\lambda_0 = 800$ nm (bandwidth of ~ 70 nm), pulse duration $T_{FWHM} = 25$ fs, repetition rate of 1 kHz, and pulse energy of 4 mJ. To ensure that all excitation laser pulses in the experiment had the same spectral bandwidth, a bandpass filter (Thorlabs FB790-10) was employed to narrow the laser bandwidth from 70 nm to 10 nm. By adjusting the laser's grating pair, the pulse chirp was set to near-zero at the front surface of the samples, resulting in nearly Fourier-transform-limited input pulses with a time duration of approximately 90 fs.

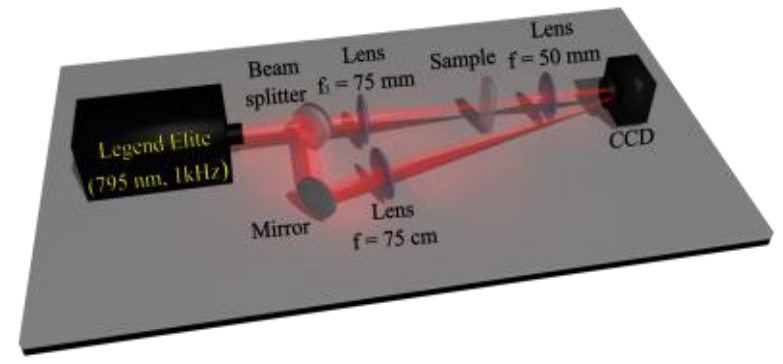

**Fig. 1:** Experimental setup.

The experimental setup is shown in Fig. 1. A beam splitter was placed before the converging lens ($f_1 = 75$ mm) to divide the laser beam into two parts: reference beam and signal beam. Both transverse intensity profiles were simultaneously captured pulse by pulse using a CCD camera (Thorlabs BC106-VIS), which was triggered by the laser.

The reference beam was focused by a lens with a focal length of $f = 75$ cm to reduce its diameter, ensuring it would fit within the region of the CCD sensor. The input signal beam was initially focused using a lens with a focal length of $f_1 = 75$ mm in two different samples:

1) Sapphire crystal with thickness of 5 mm, with its first surface positioned at the beam waist ($w_0 = 25$ μm). Subsequently, a second lens with a focal length of 50 mm was placed approximately 50 mm from the exit face of the crystal. This lens projected the light emerging from the sapphire crystal onto the CCD camera, positioned about 30 cm away from the second lens. This setup allowed the CCD to capture the projection of the light from the exit plane of the sapphire crystal.

2) Distilled water inside a cubic cuvette with thickness of 30 mm. The cuvette was positioned so that the beam waist was at its center (15 mm from both the entrance and exit faces), thus avoiding nonlinear interactions with the fused silica faces. A second lens was then placed such that its focus ($f = 50$ mm) was inside the cuvette, 5 mm from its exit face. Consequently, the second lens was positioned approximately 45 mm from the exit face of the cuvette.

To evaluate the filamentation regime in sapphire crystal, input pulses with peak powers ranging from 0.3 MW to 10.0 MW were considered. In the case of distilled water, input peak powers between 0.5 MW and 22.2 MW were considered. At 20.0 MW, multifilamentation events start to be observed sporadically in distilled water.

In Fig.2(a,b) the spectra are shown as the input power increases for sapphire and distilled water, respectively. The power thresholds required to observe supercontinuum spectra, which also correspond to the thresholds for filamentation in transparent media [4], are approximately $P_{th}^{sapphire} = 3.6$ MW and $P_{th}^{water} = 5.5$ MW. These thresholds result in broadband spectra spanning from 545 nm to 855 nm for sapphire crystal and from 470 nm to 870 nm for distilled water. At these power levels, the spectra and the beam's transverse profiles exhibit strong fluctuations, resembling turbulent behavior [24, 32-34].

Since both media considered are in the normal dispersion regime ($k'' > 0$) with self-focusing nonlinearity ($n_2 > 0$), the supercontinuum generation along the beam propagation saturates due to the temporal broadening, as a consequence, also no more spectral broadening is observed. We recall that the supercontinuum broadening not only smoothes the temporal derivatives of the pulse profile but also reduces the peak power. The saturated spectra observed for sapphire and distilled water span from 440 nm to 920 nm ($1.5P_{th}^{sapphire}$) and 380 nm to 940 nm ($1.8P_{th}^{water}$), respectively. We refer to this regime of spectral saturation as the "saturated filament".

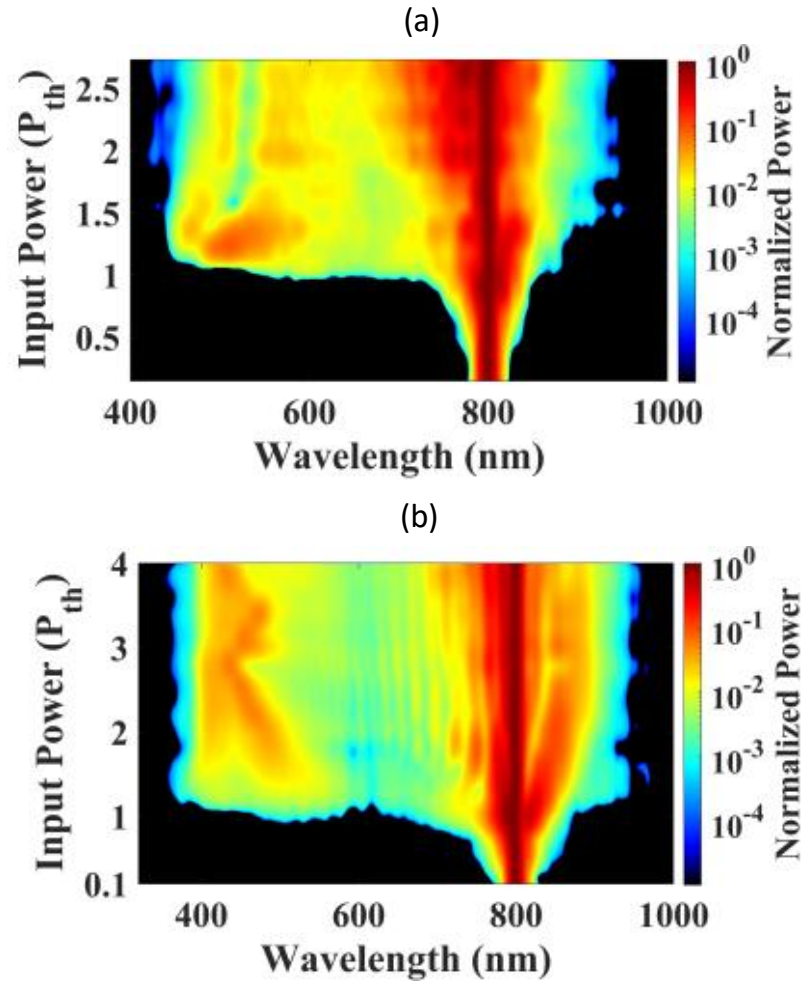

**Fig. 2:** Spectrum of signal beam as function of average input power for: (a) Sapphire crystal. (b) Distilled water.

Above the threshold power, filaments tend to be more stable since the beam reaches filamentation with more energy in its reservoir, allowing for more cycles of self-focusing (Kerr nonlinearity) and plasma defocusing. This stable filament state can be easily observed, occurring between $1.3P_{th}$ - $1.8P_{th}$ in both cases. Within these input power ranges, around $1.5P_{th}^{sapphire}$ (sapphire) and $1.4P_{th}^{water}$ (distilled water), conical emissions are observed. This process can generally be interpreted as energy scattering from the filament in other directions. Figures 3(a) and 3(b) show the typical transverse intensity profiles captured by the CCD camera for the cases of sapphire crystal and distilled water, respectively, for some input powers.

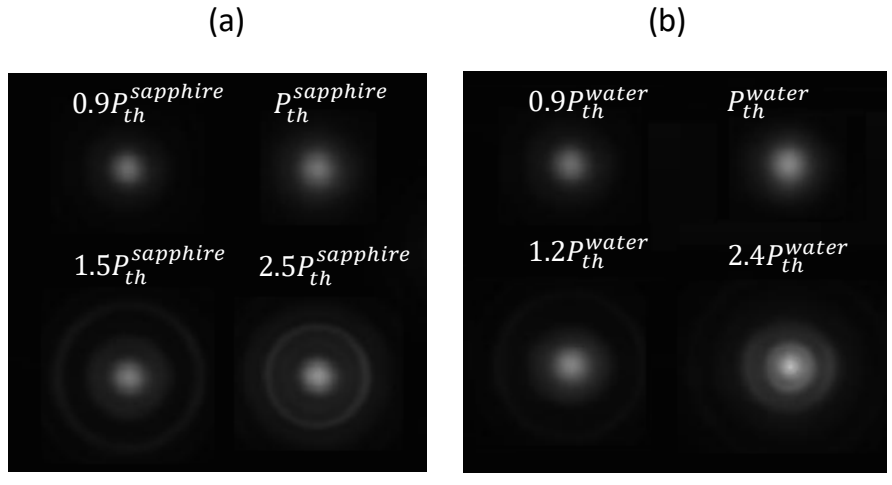

**Fig. 3:** (a,b) Examples of replicas considered for analysis of the Parisi's probability distributions for specific input powers: (a) Sapphire crystal and (b) Distilled water.

For a more consistent analysis of Parisi's distribution of the signal beam, we consider only those replicas of the reference beam that had transverse intensity profiles with a correlation coefficient of 0.99 or higher relative to the average replica and a maximum energy fluctuation of 2%. Based on this criterion, 1000 replicas of the reference pulse, along with their respective signal pulses, were selected for analysis. The Parisi distributions consistently exhibited replica-symmetric behavior, meaning the probability distribution $P(q)$ was centered at $q = 0$ in all cases for both the reference beam and the signal beam at low input powers (below the filamentation threshold), thus representing the typical behavior of passive mode-locked lasers, analogous to the ferromagnetic-like state in magnetism, where the spins are aligned in the same direction without the occurrence of frustration events [54].

As shown in Figs. 4(a) and 4(b) for sapphire crystal and distilled water, respectively, the optical transition to filamentation is marked by replica symmetry breaking, indicated by bimodality in Parisi's distribution. This behavior is similar in both the position (blue curve) and Fourier (black curve) domains of the pulse's transverse profile. Within a narrow range of input powers, slightly above the threshold power, up to approximately $P_{th}$ - $1.3P_{th}$, strong fluctuations caused by a few focusing-defocusing cycles were observed, characterized by RSB. This indicates that the attempt at self-organization during filamentation generates pulse replicas with correlated and anti-correlated fluctuations due to frustration among modes, a characteristic of photonic glassy-like systems.

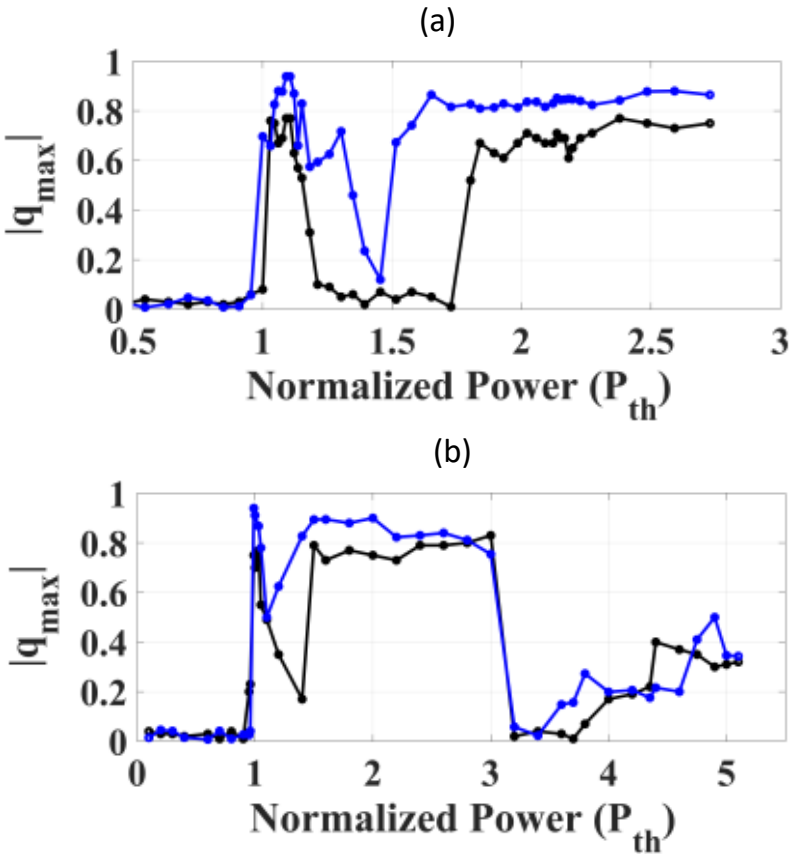

**Fig. 4:** (a,b) Loci $|q_{max}|$ of the maximum $P(q)$ versus input power for sapphire crystal (c) and distilled water (d), showing the replica-symmetric-to-RSB phase transitions in the position (black curve) and Fourier (blue curve) domains of the transverse intensity profiles: self-focusing to filamentation, saturated filament to filament with CE, and multifilamentation (only for distilled water).

As mentioned earlier, for low input power configurations, the replicas fluctuate without correlation between them ($q_{max} = 0$). For example, Figures 5(a) and 5(b) show the Parisi probability distributions for sapphire crystal and distilled water, respectively, at input powers around $\approx 0.9P_{th}$. The left and right sides represent the analysis in the position ($P_1(q)$) and Fourier ($P_2(q)$) domains, respectively. The pre-filamentation regime for both media shows that the self-focusing effect was not capable of altering the ferromagnetic-like regime of the mode-locked laser, consistently exhibiting replica-symmetric statistics. Only when supercontinuum generation and filamentation occur do interesting scenarios emerge. In these cases, a certain degree of disorder arises in the interaction due to the generated plasma, and in the case of distilled water, there is additional structural disorder due to its liquid nature.

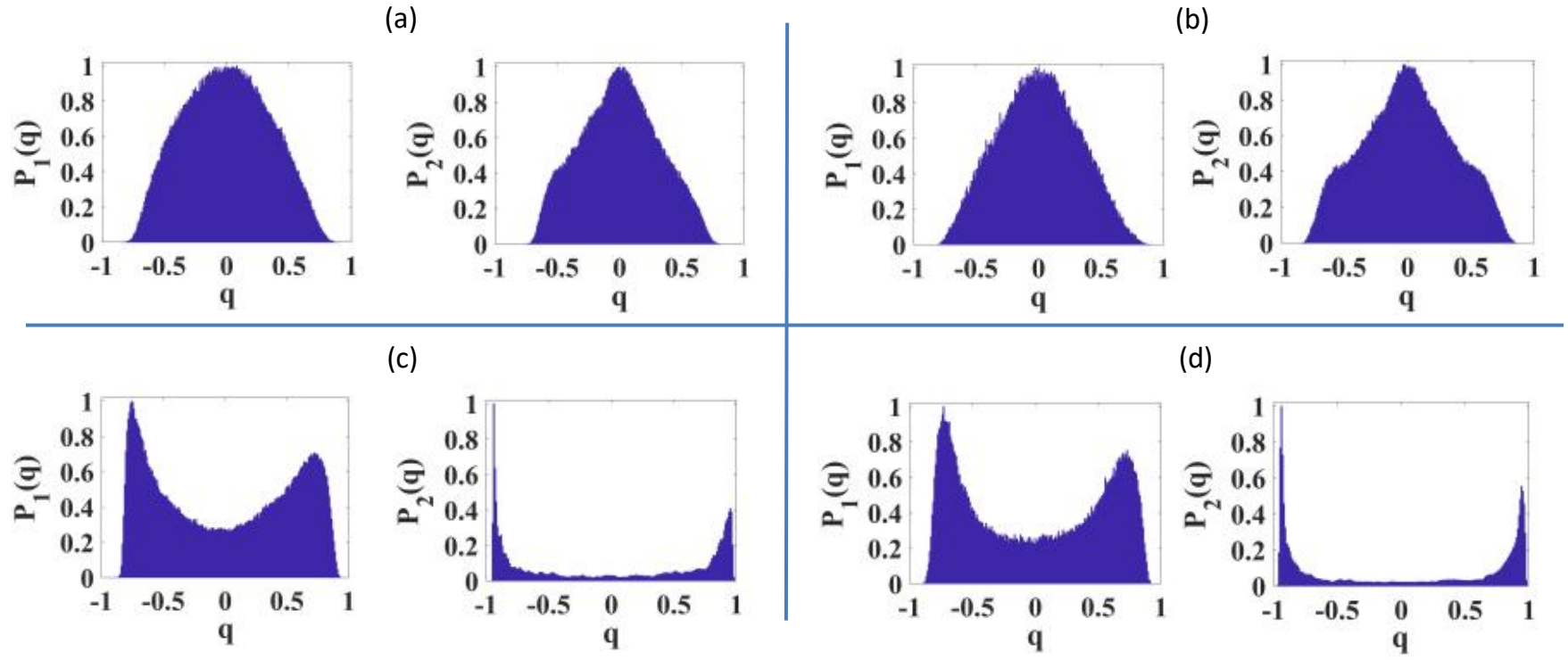

**Fig. 5:** (a-d) Parisi's probability distributions analyzing the position (left, $P_1(q)$) and Fourier (right, $P_2(q)$) domains of the transversal intensity profile of the signal beam. (a,b) Pre-filamentation regime in sapphire crystal ($0.9P_{th}^{sapphire}$) and distilled water ($0.9P_{th}^{water}$), respectively. (c,d) Near the filamentation threshold in sapphire crystal ($1.1P_{th}^{sapphire}$) and distilled water ($P_{th}^{water}$), respectively.

Figures 5(c) and 5(d) show the PPDs in the position (left side) and Fourier (right side) domains for sapphire and distilled water, respectively, for cases around $P_{th}$. Specifically, we observe a more pronounced bimodality for filamentation in sapphire at $1.1P_{th}^{sapphire}$ and in water at $P_{th}^{water}$. Our results indicate that the optical transition from ferromagnetic-like state to glassy-like state is more abrupt for water (compare the growth of $|q_{max}|$ between Figures 4(a,b)).

Interestingly, both media showed a tendency for a decrease in the bimodality of Parisi's distribution in the filament stability region (approximately $1.3P_{th}$ - $1.8P_{th}$), as indicated by the decrease in $|q_{max}|$ in Figures 4(a,b). This power range also corresponds to the saturation region in supercontinuum generation, suggesting that a saturated filament might exhibit a statistical regime with characteristics of both ferromagnetic-like and glassy-like states, similar to a random bond ferromagnet [53]. In the case of sapphire, a larger filament stability region with predominantly ferromagnetic-like behavior was observed in the position domain, while for water, its structural disorder seems to contribute to the persistence of the photonic glassy-like state, as indicated by a clearer detection of RSB (maintenance of bimodality in $P(q)$).

Within the range of input powers that provide stabler filaments, particularly from $1.45P_{th}^{sapphire}$ for sapphire and $1.20P_{th}^{water}$ for water, the first visual signs of conical emission generation appear on the CCD camera. Figures 6(a) and 6(b) show the $P(q)$ distribution for these configurations, respectively, demonstrating that in these regions we have intermediate states between ferromagnetic-like and glassy-like regimes.

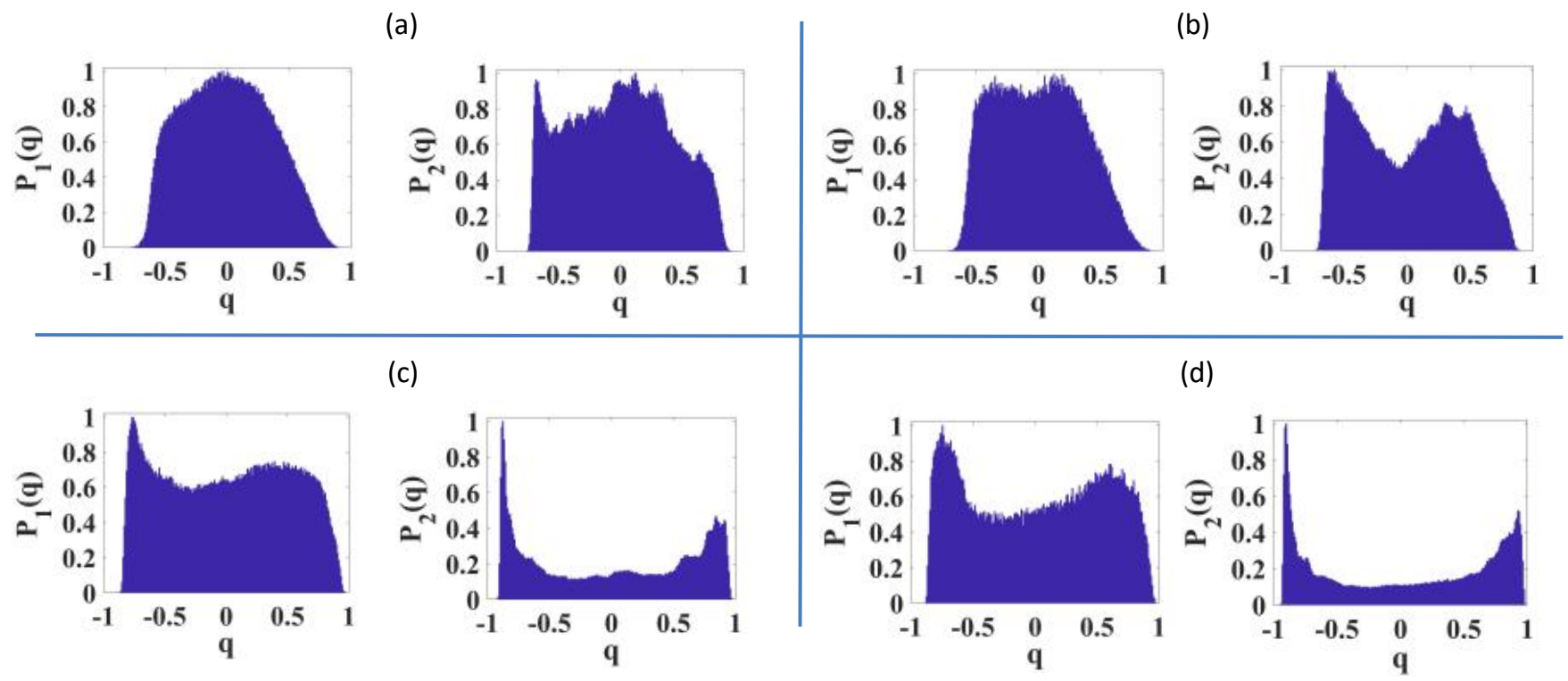

**Fig. 6:** (a-d) Parisi's probability distributions analyzing the position (left, $P_1(q)$) and Fourier (right, $P_2(q)$) domains of the transverse intensity profile of the signal beam. (a,b) Saturated filamentation in sapphire crystal ($1.45P_{th}^{sapphire}$) and distilled water ($1.2P_{th}^{water}$), respectively. (c,d) Filamentation accompanied by conical emission in sapphire crystal ($2.5P_{th}^{sapphire}$) and distilled water ($2.0P_{th}^{water}$), respectively.

However, at higher input powers, the conical emissions become more pronounced, exhibiting breathing-like fluctuations in their diameters for both propagation media. At this stage, a reversal in the behavior of $|q_{max}|$ is observed (see Figs. 4(a) and 4(b)) in both Fourier and position domains. This demonstrates that the statistical regime of the dynamics of conical emissions, along with the central filament, is also glassy-like. This is evidenced in Figs. 6(c) and 6(d), which show the PPD for sapphire and distilled water, respectively, at input powers of $2.5P_{th}^{sapphire}$ and $2.0P_{th}^{water}$. Again, we attribute the tendency of water to exhibit greater bimodality (larger $|q_{max}|$) in the PPD to its structural disorder, which contributes to more events of frustration among modes.

Note that, in general, a greater sensitivity in detecting RSB was observed through the Fourier domain of the images (see Figs. 5 and 6). Despite this, for both transparent bulk (sapphire crystal) and liquid (distilled water) media, the manifestation of RSB was consistently observed, indicating that filamentation, along with its conical emission, exhibits universal statistical regimes with photonic glassy-like behavior.

Regarding the phenomenon of multifilamentation, although its glassy-like behavior has been reported before [39], the Replica Symmetry Breaking has not yet been demonstrated. To analyze Parisi's probability distribution of multifilamentation, we considered input powers higher than $4.0P_{th}^{water}$ for distilled water. Interestingly, in the region where the input power is close to the threshold for multifilamentation, a significant decrease in the bimodality of $P(q)$ is observed, as shown in Fig. 4(b). We attribute this behavior to the loss of correlation and anti-correlation between replicas due to multiple competing events in a scenario where numerous phenomena are occurring simultaneously: filamentation, conical emission, and multifilamentation.

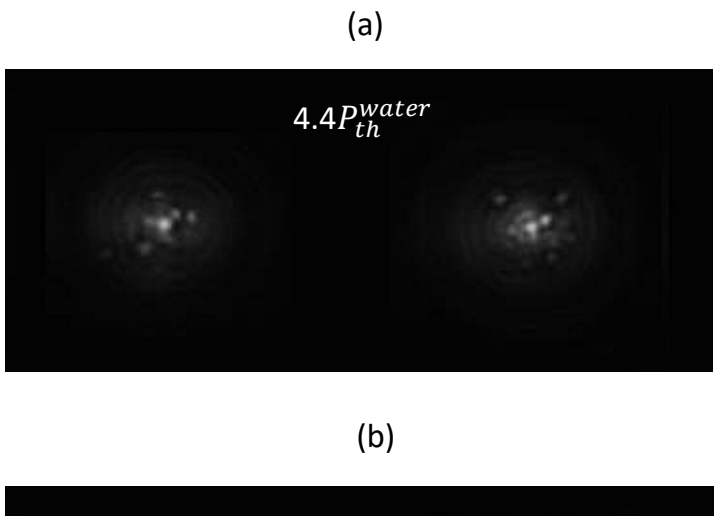

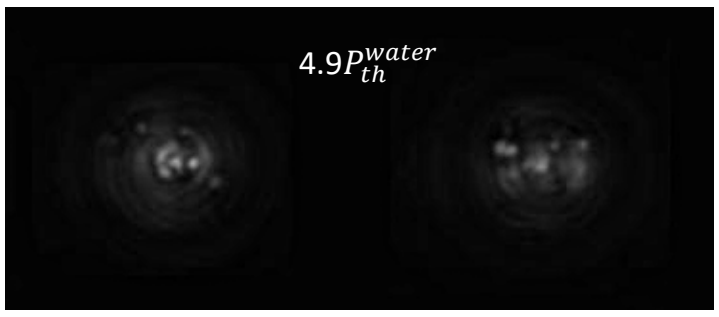

**Fig. 7:** Two representative replicas of multifilamentation in distilled water for input powers of (a) $4.4P_{th}^{water}$ and (b) $4.9P_{th}^{water}$.

Fig. 7(a) present two replicas of multifilamentation for the case that maximizes bimodality (glassy behavior) in the position domain, with an input power of $4.4P_{th}^{water}$. In Fig. 7(b), we show two replicas for the configuration that maximizes glassy behavior in the Fourier domain, corresponding to an input power of $4.9P_{th}^{water}$. The Parisi probability distributions for both cases shown in Figs. 7(a,b) are displayed in Figs. 8(a,b), respectively. In other words, Figs. 8(a,b) analyze the spatial and Fourier domains, respectively, highlighting the glassy behavior of multifilamentation most effectively.

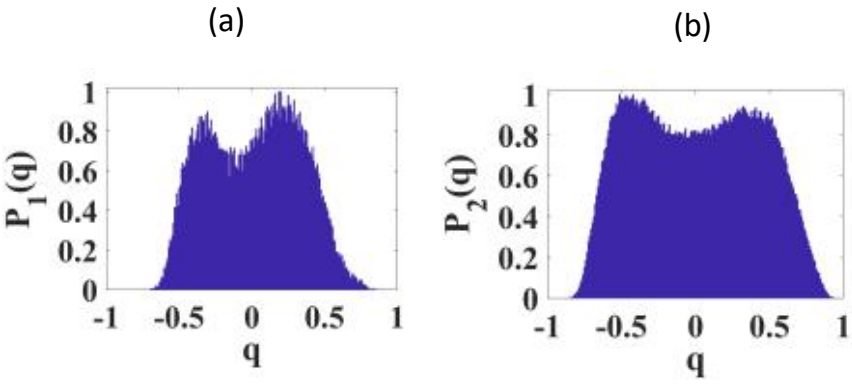

**Fig. 8:** (a) Parisi's probability distribution analyzing the position domain ($P_1(q)$) of the transverse intensity profile of multifilamentation in distilled water with an input power of $4.4P_{th}^{water}$. (b) Parisi's probability distribution analyzing the Fourier domain ($P_2(q)$) of the transverse intensity profile of multifilamentation in distilled water with an input power of $4.9P_{th}^{water}$.

Therefore, in agreement with [39], the disorder associated with nucleation (filament formation) and their interactions, well-documented in the literature, plays a key role in shaping the complex, glassy-like photonic state observed in the transverse intensity profile of multifilamentation. To better observe the glassy state of multifilamentation, we emphasize that it may be important to consider challenging strategies to mitigate the emergence of other nonlinear phenomena that could influence the PPD analysis, such as conical emission.

## 4 Conclusion

Under the framework of Parisi’s probability distribution, we experimentally investigated the statistical regimes of filamentation accompanied by supercontinuum generation in both sapphire crystal and distilled water using femtosecond pulses. Our analysis of the transverse intensity profiles and the two-dimensional Fourier transforms of beam images consistently revealed replica symmetry breaking across these domains. This observation confirms the existence of glassy-like photonic states in the formation of single filaments, filamentation with conical emission, and multifilamentation. These findings enhance our understanding of statistical nonlinear optics and provide new insights into the connections between nonlinear optics and magnetism, especially in the context of phase transitions.

## 5 Acknowledgements

We acknowledge the financial support of the Brazilian Agencies: Conselho Nacional de Desenvolvimento Científico e Tecnológico (CNPq), the Coordenação de Aperfeiçoamento de Pessoal de Nível Superior (CAPES), and the Fundação de Amparo à Ciência e Tecnologia do Estado de Pernambuco (FACEPE). The postdoctoral fellowship of A.C.A.S. and G.P. were provided by Office of Naval Research (ONR).